\documentclass{webofc}

\usepackage[varg]{txfonts}   
\usepackage{hyperref}
\usepackage{url}
\usepackage{graphicx}
\usepackage{amsmath,amssymb}
\hypersetup{colorlinks=true,citecolor=blue,urlcolor=blue,linkcolor=blue}
\begin{document}
\title{Deformation of Jets Induced by Ambient Medium Flow}
%
%

\author{\firstname{Arjun} \lastname{Sengupta}\inst{1,2}
\fnsep\thanks{\email{aseng@tamu.edu}} 
        \and
        \firstname{Rainer J.} \lastname{Fries}\inst{1,2}\fnsep\thanks{\email{rjfries@tamu.edu}} }

\institute{Cyclotron Institute, Texas A\&M University, College Station TX 77843, USA
\and
           Department of Physics and Astronomy, Texas A\&M University, College Station TX 77843, USA
          }

\abstract{The evolution of jets showers in high energy nuclear collisions is influenced in various ways by the presence of a surrounding medium. The interaction of jet constituents with the medium can happen during the partonic stage of the jet, during hadronization, and even during its hadronic stage. We demonstrate how flow of the ambient medium in a direction transverse to the jet can introduce both dipole and quadrupole defomations. We propose to analyze the $n=1$ and $n=2$ harmonic deformations of soft and semi-hard hadrons or subjets in a jet with respect to the jet core using the method of $q$-vectors. We discuss simulations which show how the transverse shapes and their preferred angles evolve when the ambient environment of jets changes from the vacuum to a parton medium without flow and finally to a medium with various rates of transverse flow. Our study includes the effects of both flow during the development of the parton shower and hadronization.
The existence of dipole deformations, and the correlation of the angles of dipole and quadrupole deformations could constitute promising experimental signals for the presence and size of ambient transverse flow.
}
\maketitle
\section{Introduction}
\label{intro}

Jets and high momentum hadrons are well established probes of quark-gluon plasma (QGP) that allow us to estimate important transport properties of the latter. However, idealizing assumptions are often still necessary in calculations, even when jets are modeled in a purely perturbative regime. For example, the ambient medium for a jet is often assumed to be homogenous and static, even though it is possible to model its dynamic evolution quite accurately by state-of-the art viscous fluid dynamics codes. 

It had been speculated since the early years of jet quenching, that flow in the ambient medium should lead to observable consequences for jets \cite{Armesto:2004pt}, but only recently has there been a broader effort to incorporate both gradient effects and flow effects into jet quenching theories \cite{He:2020iow,Sadofyev:2021ohn,Antiporda:2021hpk,Barata:2023zqg}. The push to incorporate flow effects will lead to more accurate theoretical descriptions. But it is also motivated by the promise of new observables that are sensitive to conditions in the medium, or perhaps even the prospect of a new type of tomography of the QGP fireball. 

In this work we will explore if transverse flow can lead to deformations of the transverse shape of jets, and explore possible ways to ascertain the existence of such deformations beyond those from typical fluctuations present even without flow. In the next section we discuss the basic ideas of this work, and in the following section we present some Monte Carlo simulations of jet ensembles which show that at least in a simplified situation clear jet shape deformations are indeed present.

\section{Transverse Flow and Jets}
\label{sec-1}

It is worthwhile to first discuss how ambient flow transverse to the propagation direction of a jet can arise. Three scenarios seem the most plausible. 

(1) The QGP fireball experiences strong \emph{global longitudinal} flow away from mid-rapidity (Bjorken expansion). Jets at forward or backward rapidity might experience a differential between their own longitudinal rapidity and the ambient rapidity, leading to an effective flow of the medium transverse to the jet direction.

(2) Jets that are created toward the edge of the fireball and propagate in a tangential direction might experience the onset of \emph{global radial} flow as a flow transverse to their direction of propagation. However, unlike in scenario (1), which sees longitudinal flow present from the beginning, radial flow builds up with time and might still be rather small when jets leave the fireball.

(3) Even more subtle, but nevertheless possible, are \emph{local transverse flow} effects from fluctuations of the fireball in the transverse plane. Plainly, a hotspot on one side of a jet trajectory accompanied by a less dense spot on the opposite side will lead to the build up of flow transverse to the jet. 

All three scenarios should be be studied by detailed 3-D fluid dynamic simulations with realistic initial conditions in the future.

\begin{figure}[h]
\centering
\includegraphics[width=12cm,clip]{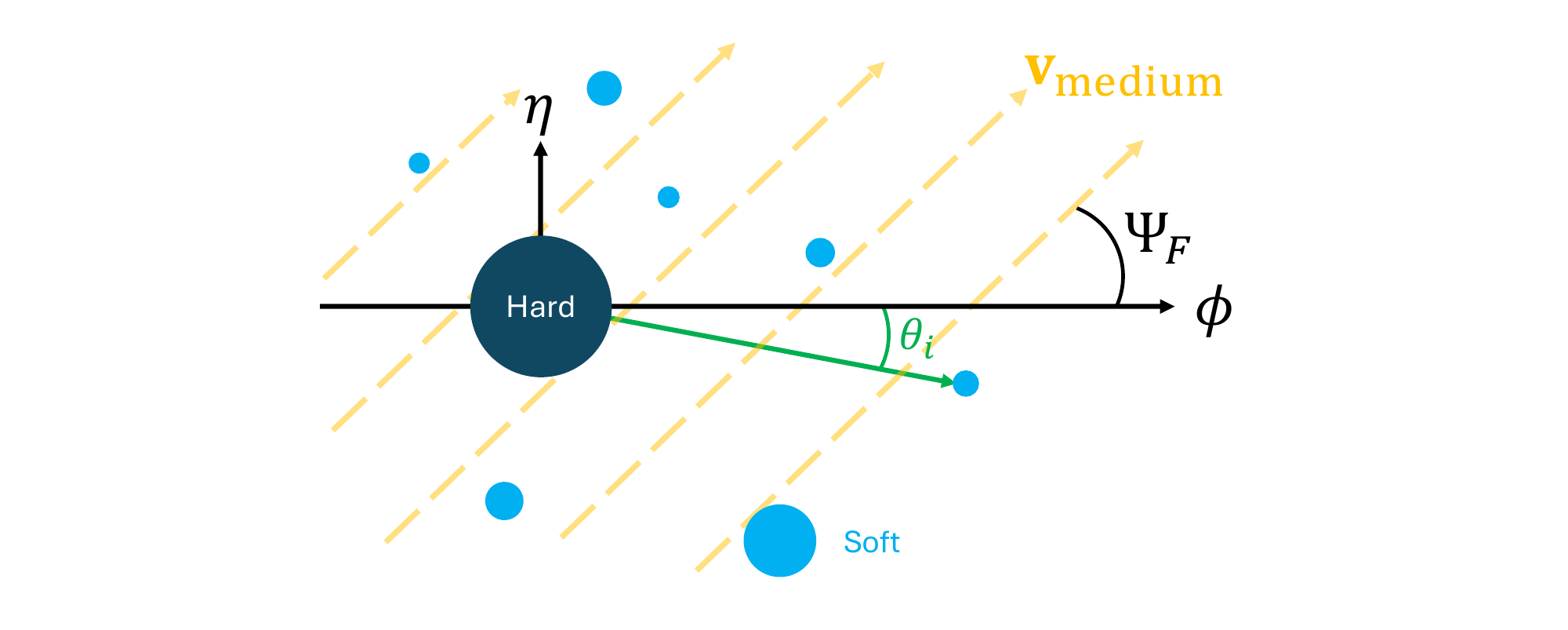}
\caption{Cartoon of a jet in the $\eta$=$\phi$ plane exhibiting a core and several "soft" objects around it seen under angles $\theta_i$. Ambient flow at a preferred angle $\psi_F$ is also shown. The question we would like to answer is whether the soft objects in the jet are sensitive to $\psi_F$.}
\label{fig-1}       
\end{figure}

Fig.\ \ref{fig-1} shows the cartoon of a jet in the $\eta$-$\phi$ plane as consisting of a hard core and several soft or semi-hard associated objects. These could be hadrons or subjets reclustered with a small radius $R$. As an example, in the simulations in the next section we will use jets reclustered with $R=0.1$ and define the core as the hardest subjet and "soft" objects as subjets with momentum between 2 and 10 GeV. In the vacuum or a medium without a preferred direction, we expect the soft objects to be randomly distributed in azimuthal angle around the core. 

However, if a preferred direction is present in the ambient medium due to medium flow, as indicated in Fig.\ \ref{fig-1}, soft subjects should be sensitive to momentum kicks in the preferred direction. These momentum kicks can happen in the partonic energy loss phase, where --- unlike in an isoptropic medium ---  the flow velocity $\mathbf{u}$ can lead to $\langle \mathbf{k}_\perp\cdot \mathbf{u}\rangle \ne 0$. It can also occur during hadronization in which shower partons can pick up flowing medium partons. In the simulations below both of these processes are modeled.

It should thus be possible to study the multipole structure
of the deformation, e.g.\ by using the method of $q$-vectors,
as has been successfully done for global flow analysis in heavy ion collisions \cite{Poskanzer:1998yz}.
We define
\begin{equation}
  \mathbf{q}_n = \frac{1}{N} \sum_{i=1}^N \begin{pmatrix} \cos(n\theta_i) \\ \sin(n\theta_i) \end{pmatrix}
\end{equation}
for integer $n$ for each jet with $N$ soft objects. The corresponding "event plane" angle at order $n$ for a jet is given by 
\begin{equation}
    \psi_n = \frac{1}{n}\arctan\frac{q_n^2}{q_n^1}  \, .
\end{equation}
We expect that kicks of jet shower partons in a preferred direction due to flow will lead to non-vanishing dipole ($n=1$) deformations. Furthermore the angle $\psi_1$ should correlate with the flow angle $\psi_F$.

\section{Simulations}

We test the ideas from the previous section using the JETSCAPE 3.0 Monte Carlo \cite{Putschke:2019yrg}. It uses MATTER for parton showering and LBT for the propagation of partons through QGP. Ambient flow in these modules is implemented through Lorentz boosts into the local fluid rest frame in which the calculations are done, followed by a boost back to the laboratory frame. Hadronization is carried out by Hybrid Hadronization \cite{Han:2016uhh,JETSCAPE:2025wjn} which allows the recombination of shower partons with sampled medium partons as well as string fragmentation. As was shown in Ref.\ \cite{JETSCAPE:2025wjn} medium flow effects can be transferred to jets through in-medium hadronization.

As a first step we simulate flow effects in a static medium in which temperature and flow conditions can be set manually (aka a QGP brick). We have run simulations with varying temperatures and medium lengths, but the results shown here are all obtained with $T=300$ MeV and a brick size of 4 fm. Jets are always initiated to travel in the $x$-direction which allows for a convenient analysis of particles and jets in the typical collider variables $\eta$ and $\phi$. Jets are given an initial momentum of 100 GeV.

\begin{figure}[h]
\centering
\includegraphics[width=13cm,clip]{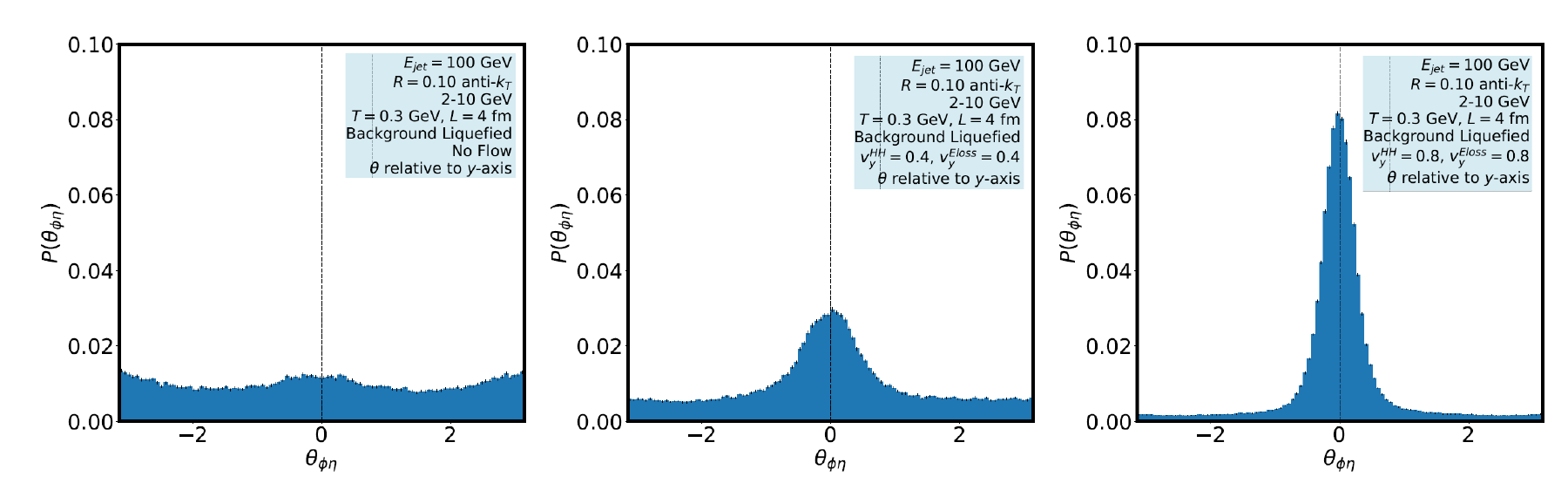}
\caption{The distribution of angles \{$\theta_i$\} of $R=0.1$  soft subjets in the $\eta$-$\phi$-plane for jets in a medium without flow (left panel), with transverse flow velocity $v=0.4$ during (center panel) and transverse flow velocity $v=0.8$ (right panel). The angles $\theta_i$ are defined with respect to the medium flow direction.}
\label{fig-2}       
\end{figure}

Motivated by Fig.\ \ref{fig-1} we first analyze the angles $\{\theta_i\}$ of soft subjets (defined to have between 2 and 10 GeV momentum). Fig.\ \ref{fig-2} shows the distribution for large samples of jets run through either a QGP medium without flow, a medium with intermediate flow (v=0.4) and a medium with very large flow (v=0.8). The flow is fixed to be in the $y$-direction for this exercise. Note that the flow is kept constant and is the same during the parton shower phase ("$v^{Eloss}$") and during hadronization ("$v^{HH}$"). It was noted in Ref.\ \cite{JETSCAPE:2025wjn} that the hard core of a jet is largely unaffected by transverse flow. However, as we see in Fig.\ \ref{fig-2}, flow introduces a clear correlation of the azimuthal angle of the soft subjets with the flow angle. It grows with the magnitude of the flow velocity.

\begin{figure}[h]
\centering
\includegraphics[width=13cm,clip]{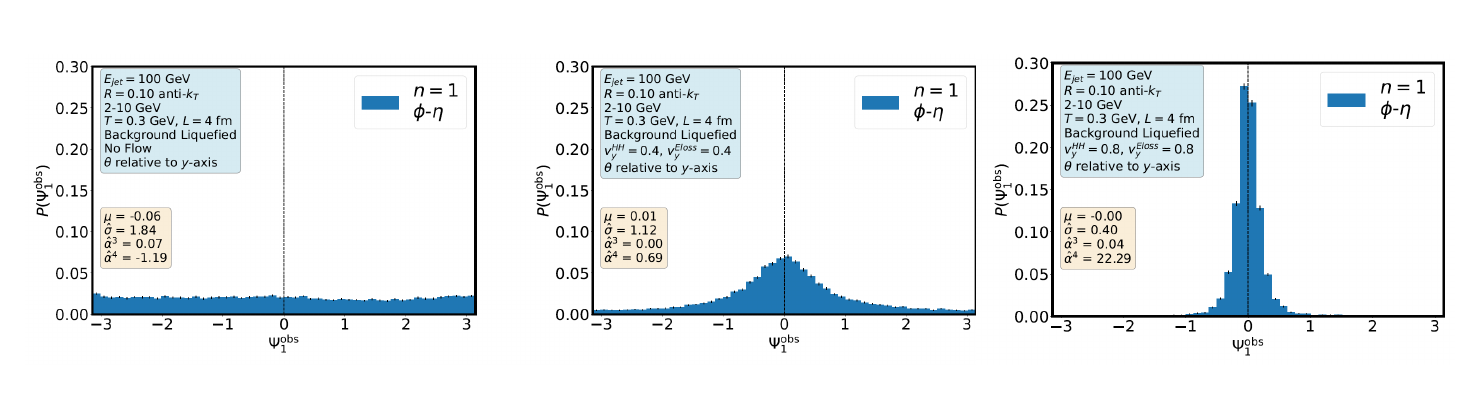}
\caption{The distribution of dipole angles \{$\psi_i$\} in the $\eta$-$\phi$-plane for jets as defined in the text. Jets without flow (left panel), with transverse flow velocity $v=0.4$ during (center panel) and transverse flow velocity $v=0.8$ (right panel) are shown. In the center and right panel the flow angle $\psi_F$ is set to be zero and therefore the observed dipole angles are highly correlated with the flow angle.}
\label{fig-3}       
\end{figure}

Fig.\ \ref{fig-3} shows the distribution of the dipole "event plane" angle $\psi_1$ for ensembles of jets again in bricks without flow and with transverse flows of magnitude 0.4 and 0.8. Already with the smaller flow value we see a strong correlation of the preferred direction of the dipole with the direction of the ambient flow. This correlations sharpens with increasing flow. 

Somewhat surprisingly, our simulations also yield a sizeable quadrupole deformation ($n=2$) for jets (not shown here). The angle $\psi_2$ also exhibits a strong correlation with the ambient flow angle, similar to the behavior of $\psi_1$.

\section{Summary}

\begin{figure}[h]
\centering
\includegraphics[width=8cm,clip]{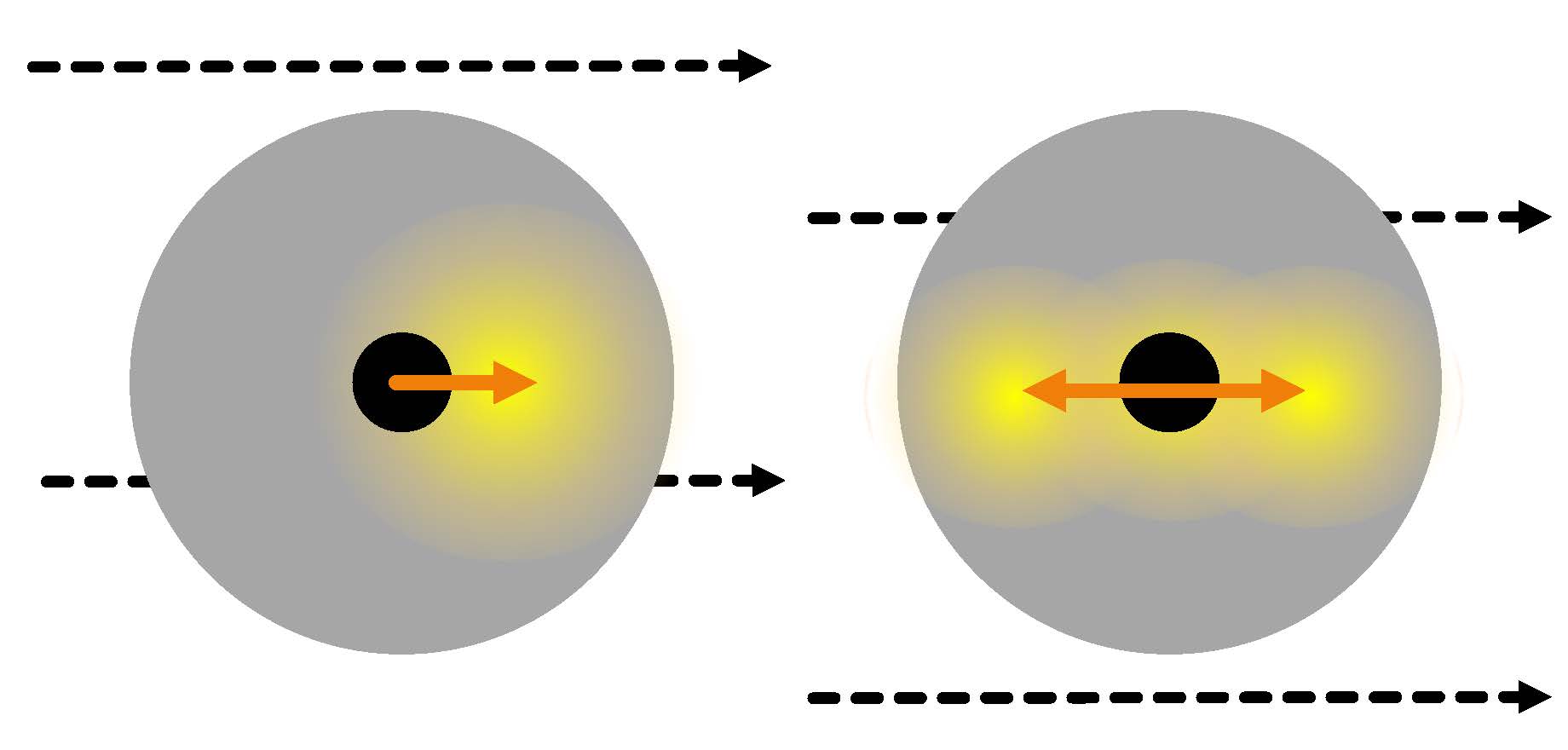}
\caption{Schematic summary of our simulations. Soft subjets (yellow) are both on average deflected away from the jet core (dipole deformation, left) and elongated (quadrupole deformation, right). Both of these deformations correlate with the ambient flow direction.}
\label{fig-4}       
\end{figure}

Fig.\ \ref{fig-4} represents the core ideas of our results. 
For sufficiently large ambient flow, jets exhibit deformations that could be analyzed jet by jet through $q$-vector analysis. However, the values for the flow velocities discussed here are certainly larger than anything one can expect in scenarios (2) and (3) discussed in Sec.\ \ref{sec-1}. Realistic deformations could therefore be subtle. One idea that should be explored is the combination of different observables. E.g., by choosing jets with matching $\psi_1$ and $\psi_2$ angles, within a certain band, one can expect to construct a jet sample enriched with jets that have experienced transverse flow.
It would also be advisable to vary other parameters of our study like the exact definition of soft objects, to optimize flow signals. Eventually, full Monte Carlo studies including realistic 3-D simulations of the soft background medium will need to be carried out.



%
%
%

This work was supported by NSF awards PHY-2111568, PHY-2413003 and OAC-2004571. This work has used resources provided by  Texas A\&M High Performance Research Computing.

\end{document}